\documentclass[twocolumn,secnumarabic,amssymb, nobibnotes, aps, prl]{revtex4-1}

\usepackage{amsmath}
\usepackage{amssymb}
\usepackage{graphicx}
\usepackage{epstopdf}
\usepackage{upgreek}

\begin{document}

\title{Nonlinear response in high-field dielectric laser accelerators}

\author{D. Cesar}
\affiliation{Department of Physics and Astronomy, UCLA, Los Angeles, California 90095, USA}

\author{S. Custodio}
\affiliation{Department of Physics and Astronomy, UCLA, Los Angeles, California 90095, USA}

\author{J. Maxson}
\affiliation{Department of Physics and Astronomy, UCLA, Los Angeles, California 90095, USA}

\author{P. Musumeci}
\affiliation{Department of Physics and Astronomy, UCLA, Los Angeles, California 90095, USA}

\author{X. Shen}
\affiliation{Department of Physics and Astronomy, UCLA, Los Angeles, California 90095, USA}

\author{E. Threlkeld}
\affiliation{Department of Physics and Astronomy, UCLA, Los Angeles, California 90095, USA}

\author{R.\,J. England}
\affiliation{SLAC National Accelerator Laboratory, Menlo Park, CA, 94025, USA}

\author{A. Hanuka}
\affiliation{SLAC National Accelerator Laboratory, Menlo Park, CA, 94025, USA}

\author{I.\,V. Makasyuk}
\affiliation{SLAC National Accelerator Laboratory, Menlo Park, CA, 94025, USA}

\author{E.\,A. Peralta}
\affiliation{SLAC National Accelerator Laboratory, Menlo Park, CA, 94025, USA}

\author{K.\,P. Wootton}
\affiliation{SLAC National Accelerator Laboratory, Menlo Park, CA, 94025, USA}

\author{Z. Wu}
\affiliation{SLAC National Accelerator Laboratory, Menlo Park, CA, 94025, USA}

\date{today}

\begin{abstract}
Laser powered dielectric structures achieve high-gradient particle acceleration by taking advantage of modern laser technology capable of producing electric fields in excess of 10~GV/m. These fields can drive the bulk dielectric beyond its linear response, and break the phase synchronicity between the accelerating field and the electrons. We show how control of the pulse dispersion can be used to compensate the effect of self-phase modulation and maximize the energy gain in the laser accelerator.In our experiment, a high brightness 8~MeV e-beam is used to probe accelerating fields of 1.8~GV/m in a “grating-reset” dielectric structure illuminated by a 45~fs laser pulse with a fluence of 0.7~J/cm$^2$.
\end{abstract}

\pacs{06.60.Jn}

\maketitle

Using intense optical fields for particle acceleration is at the focus of much research in modern accelerator and beam physics. Dielectric laser accelerators (DLAs)~\cite{england_dielectric_2014} are photonic structures which mediate the transfer of energy from a laser to electrons by overcoming the dephasing inherent to laser-electron interactions in vacuum~\cite{lawson_lasers_1979}. The electron dynamics in a DLA are uniquely influenced by both amplitude and phase of the drive laser wave, leading to a rich environment for studying both material physics and accelerator design.


An attractive feature of DLAs is the miniaturization associated with their orders of magnitude smaller drive wavelength as compared to conventional rf-based accelerators. Nano-fabrication techniques~\cite{simakov_diamond_2017, peralta_design_2012, ceballos_fabrication_2015} have driven the introduction of novel accelerating, focusing, and diagnostic structures~\cite{wootton_dielectric_2016, rosenzweig_galaxie_2012, bar-lev_plasmonic_2014}. At the same time, very large (multi-GV/m) fields are achievable using commercial laser systems, thus offering the potential for widely available, compact high energy accelerators for a variety of applications in science and industry~\cite{mourou_future_2013}.

Many DLA designs are based on a sub-wavelength diffraction grating which accelerates electrons by scattering a fraction of the incoming laser into a mode with Bloch-type periodicity imposed by the grating $\lambda_g = 2\pi / k_g$. The scattered mode can resonantly interact with electrons of velocity $\beta_z$ if the grating geometry is chosen to satisfy the phase matching condition (i.e. $k_g - \omega / \beta_z c \approx 0 $). This technique for controlling the fields at the optical scale has been experimentally demonstrated using relativistic and non-relativistic resonant velocities, at SLAC~\cite{peralta_demonstration_2013,wootton_demonstration_2016}, FAU Erlangen~\cite{breuer_laser-based_2013, mcneur_miniaturized_2016}, and Stanford~\cite{leedle_dielectric_2015, leedle_laser_2015}.

Maintaining the phase-matching condition over multiple cycles requires precise control of the laser envelope. For short accelerating sections (small energy gains) this means preparing the drive laser with a flat phase profile in order to keep the accelerating wave synchronous with electrons of a given energy. But as the incident electric field is increased, the material begins to exceed the linear response of the dielectric structure and gives rise to a variety of interesting phenomena including self-phase modulation, self-focusing and induced free carrier density~\cite{agrawal_chapter_2013-1,schiffrin_optical-field-induced_2012,sudrie_femtosecond_2002, hoyo_rapid_2015}, which can distort the laser amplitude and phase profiles. In particular, the optical Kerr effect can accumulate in the dielectric substrate of the grating and cause electrons to dephase from the drive laser, hindering significant acceleration~\cite{koyama_parameter_2014}. The study of soliton propagation in nonlinear optics~\cite{agrawal_chapter_2013-1} suggests that negative dispersion can effectively compensate this dephasing by providing an opposite phase curvature to cancel the effects of nonlinear propagation.

\begin{figure}[b]
\centering
\includegraphics[width=.45\textwidth]{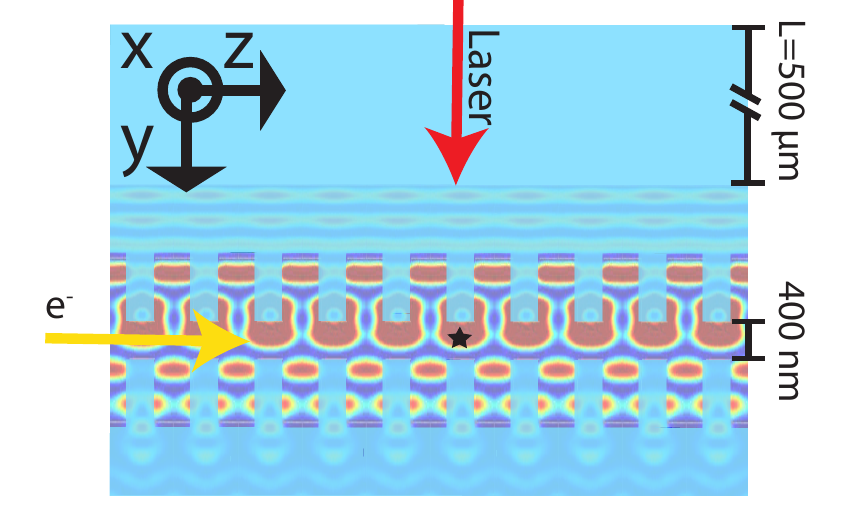}
\caption{Cartoon illustrating the fields in a dual-grating fused silica DLA. The electrons (yellow arrow) propagate from left to right through a vacuum gap between the gratings while the laser (red arrow) travels from top to bottom before being scattered by the grating layer. The star indicates the location in a unit cell for which Fig.~\ref{fig:simFROG} illustrates $\mathcal{E}(t)$.}
\label{fig:setup}
\end{figure}

In this paper we explore high field phenomena in dielectric laser acceleration by externally injecting a high brightness 8~MeV electron beam from the UCLA Pegasus photoinjector~\cite{maxson_direct_2017} to probe the GV/m accelerating fields in the gap of a double-grating fused silica dielectric structure. For intensities exceeding 7~TW/cm$^2$, the electron energy spectra indicate that significant dephasing occurs due to nonlinear propagation of the drive laser pulse through the bulk fused silica. We also demonstrate the effectiveness of a pre-compensation scheme based on adding negative dispersion to the laser pulse to counteract the optical Kerr effect and maximize the DLA-induced energy modulation. Our observations are in excellent agreement with both time and frequency-domain calculations and demonstrate that the peak accelerating field in the vacuum gap of the DLA reaches 1.8~GV/m.

A cartoon depicting the electron based measurement of the accelerating fields in a DLA is illustrated in Fig.~\ref{fig:setup}. The accelerating structure consists of two gratings made by etching teeth 700~nm tall by 325~nm wide with 800~nm periodicity into 500~$\mu$m thick fused silica wafers. The wafers are then bonded together to leave either a 400 or 800~nm vacuum channel~\cite{peralta_accelerator_2015, plettner_proposed_2006}. A round 300~fC electron beam of $\gamma_0 m c^2$ = 8~MeV energy, incident from the left, is focused (with rms dimensions $\sigma_{(x,y)}$$\sim$10~$\mu$m, $\sigma_{{(x',y')}}$$\sim$1~mrad, $\sigma_z$$\sim$1 ps) into the vacuum channel, whose small phase space aperture (hard edge 400~nm $\times$ 0.5-1~mrad) permits only 1-2\% of the incident electron beam to be propagated downstream. Using a fluorescent screen, the bunch is spatially overlapped with a Ti:Sapphire laser pulse (1/e$^2$ $w_x=45~\mu$m and $w_z=500~\mu$m, $\tau_\text{FWHM}$$\sim$45fs) which is propagating in $\hat{y}$, perpendicular to the electrons, and polarized along $\hat{z}$, the electron-beam axis. Relative time-of-arrival is first found by using the drive laser to ablate a copper grid located near the DLA , rapidly ($<1$~ps) creating a long-lived ($\sim$15~ps) electron-gas which can distort a point-projection image of the grid formed by focusing the electron beam before the grid~\cite{scoby_effect_2013}; time-of-arrival is then refined using the laser acceleration signal itself. The result of the interaction in the DLA is then assessed by measuring the electron's energy distribution with a magnetic spectrometer.

\begin{figure}[]
\centering
\includegraphics[width=.45\textwidth]{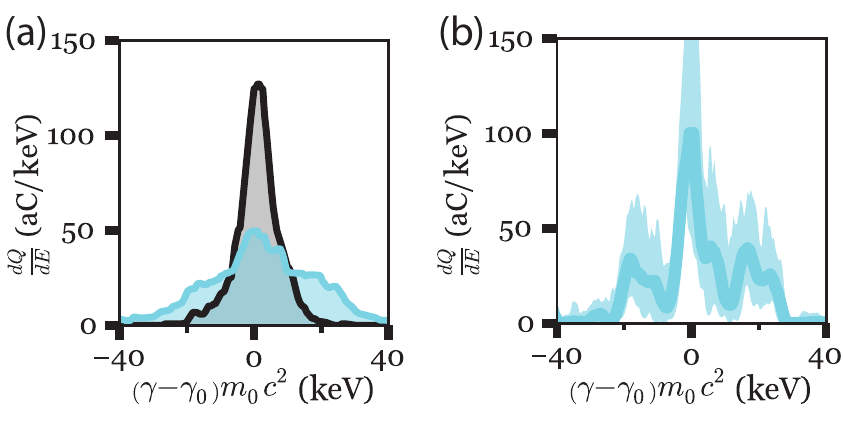}
\caption{(a) An example electron energy distribution at the spectrometer screen for typical laser-on(blue) and laser-off(black) shots. Both spectra contain a total charge of nearly 3~fC. (b) Deconvolution of the two spectra on the left. The shaded region bounds the variation caused by deconvolving the on-shot on the left with many independent off-shots.}
\label{fig:electronspectra}
\end{figure}

The laser-induced energy spread, obtained from the experimental energy spectra (an example is shown in Fig.~\ref{fig:electronspectra}), is a direct measurement of the integrated accelerating field, and can be used to infer the structure efficiency if the input fields are known. Assuming the laser pulse has many cycles, but is much shorter than the grating structure, neglecting the small $y$ dependence of fields in the vacuum gap, and using kinematic electron trajectories, we have:
\begin{equation}\label{eq:1}
\Delta E=\left| \int_{-\infty}^{\infty}\alpha E_0 \mathcal{E}(\vec{r}=\vec{r_0}+\vec{\beta}ct,t)dz \right| \cos(\omega t_0)
\end{equation}
where $\omega t_0$ is the electron's injection phase, $\mathcal{E}(\vec{r},t)$ is the complex envelope of the $E_z$ field incident upon the grating layer, $\vec{r_0}$ is the initial position of the particles and $\alpha$ gives the fraction of the incident electric field amplitude $E_0=\sqrt{2Z_0 I}$ which is scattered into the accelerating mode: $\frac{\partial (\gamma m c^2)}{\partial z}=\alpha E_0 \mathcal{E}$. Measuring the electron energy spectra for a variety of incident field strengths ($E_0$) and envelopes ($\mathcal{E}\left(\vec{r},t\right)$) allows us to characterize the performance of the accelerator.

All of the spectra exhibit a characteristic symmetric broadening due to the uniform distribution of the initial phases $\omega t_0$ for the 1~ps~(rms) long electron bunch. The resolution of the magnetic spectrometer, $\sim$3~keV~(rms), is the main contribution to the width in the laser-off shots and is limited by the transverse size of the electron beam and the point spread function of the YAG:Ce scintillating screen and its camera. Because of this, the laser-off shots all have similar profiles (when normalized by total charge), and we can determine the energy modulation by de-convolving separate laser-on and laser-off distributions, as in Fig.~\ref{fig:electronspectra}(b), in order to retrieve the effect of the laser acceleration. The deconvolution clearly shows charge at three locations: the two peaks at $\pm$ 18.5 keV are electrons at the phases corresponding to peak acceleration and to peak deceleration, while the peak at zero is caused by electrons which did not overlap with the laser field. From the deconvolution we define a robust measure of the maximum energy gain as half of the energy spread which includes 75 percent of the charge.

\begin{figure}[]
\centering
\includegraphics[width=.45\textwidth]{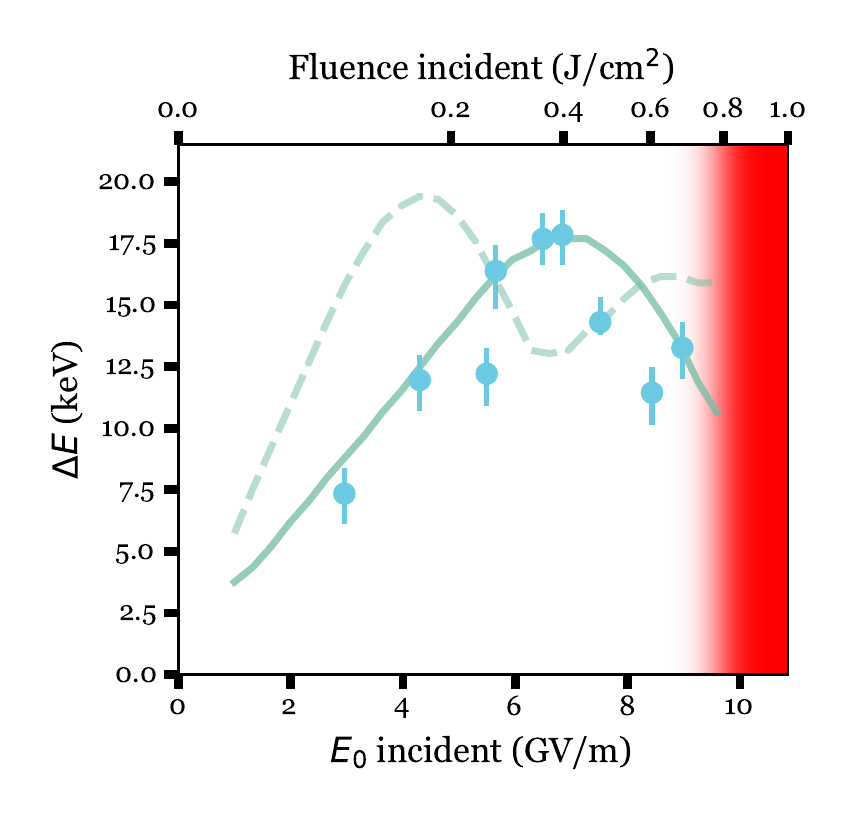}
\caption{Energy gain versus amplitude of the incident field. The red region indicates the damage fluence. The indicated uncertainty accounts for variation in both the laser-on and laser-off shots. The dashed green line shows the simulated energy gain for a perfectly-aligned electron beam and the solid green line shows a sample curve for an electron beam misaligned within the experimental uncertainty. The saturation is caused by dephasing, but the amplitude of the electric field in the mode resonant with the electrons increases linearly to 1.8~GV/m}
\label{fig:gradvse0}
\end{figure}

We demonstrate the processes underlying high field acceleration by extracting the electron energy gain as a function of the incident pulse energy (Fig.~\ref{fig:gradvse0}) in a 400~nm gap DLA. The drive laser is first compressed by maximizing the second harmonic generation from a thin nonlinear crystal located in the same plane as the DLA, and then the pulse energy is converted to $E_0$ using the laser transverse and longitudinal profiles, as measured by a camera and frequency resolved optical gating (FROG)~\cite{trebino_frequency-resolved_2000}. For incident fields below 7~GV/m the energy gain increases linearly, as expected from Eq.~\ref{eq:1} and in agreement with previous measurements~\cite{peralta_accelerator_2015, wootton_demonstration_2016}. But above 7~GV/m the energy gain rapidly saturates in a fully reversible process, which we attribute to a rapid change in $\mathcal{E}$ caused by the nonlinear Kerr effect.

Before the high-intensity laser reaches the grating layer, it propagates through a 500~$\mu$m fused silica wafer where it excites a third order nonlinear polarization field which modulates the accelerating field probed by the electron beam. The primary effect of the nonlinear response is an intensity-dependent phase modulation: $\Delta\Phi \approx n_2 I k_0 L$, with $n_2$=$2.48\cdot10^{-16}$~cm$^2$/W~\cite{milam_review_1998} consistent with an independently measured z-scan~\cite{sheik-bahae_sensitive_1990} through the substrate. In a simplified picture, this self-phase modulation works to saturate the energy gain by forcing an (otherwise synchronous) electron to sample a changing phase. For an incident electric field of 5 GV/m, the peak phase change is nearly $\pi$ radians, causing the sign of the field to flip, effectively halting the acceleration. This description suggests a clear method for compensating the saturation: use a grating pair to prepare a laser having chirp with the opposite curvature of the nonlinear phase, and thus accelerate the electrons with a flat phase.

A numerical simulation of the nonlinear DLA response, discussed later and in the supplemental material, yields a prediction we can compare to the measured energy gain. We simulate two cases, shown as green curves in Fig.~\ref{fig:gradvse0}: the first uses a single electron and perfect alignment between the laser, electron, and DLA (dashed green line); while the second includes a realistic electron beam distribution($\sigma_x=\sigma_y=10$~$\mu$m) and a sample misalignment ($x=10$~$\mu$m, $y'=5$~mrad) within the experimental tolerances. In both cases $\Delta E$ rises linearly before saturating, but because the drive laser is narrowly focused, some electrons in the realistic beam see a lower field and saturate at a higher (incident) $E_0$. Once most individual trajectories have reached saturation, the $\Delta E$ plateaus at a level set only by $\alpha$ and $n_2$, independent of misalignments, from which we obtain $\alpha = 0.2 \pm 0.04$. Since the saturation is fully explained by dephasing, we can infer that the magnitude of the longitudinal field in the mode resonant with the electrons is $\alpha E_0$, or 1.8$\pm$0.3~GV/m for the highest incident field measured in Fig.\ref{fig:gradvse0}.

\begin{figure}[]
\centering
\includegraphics[width=.45\textwidth]{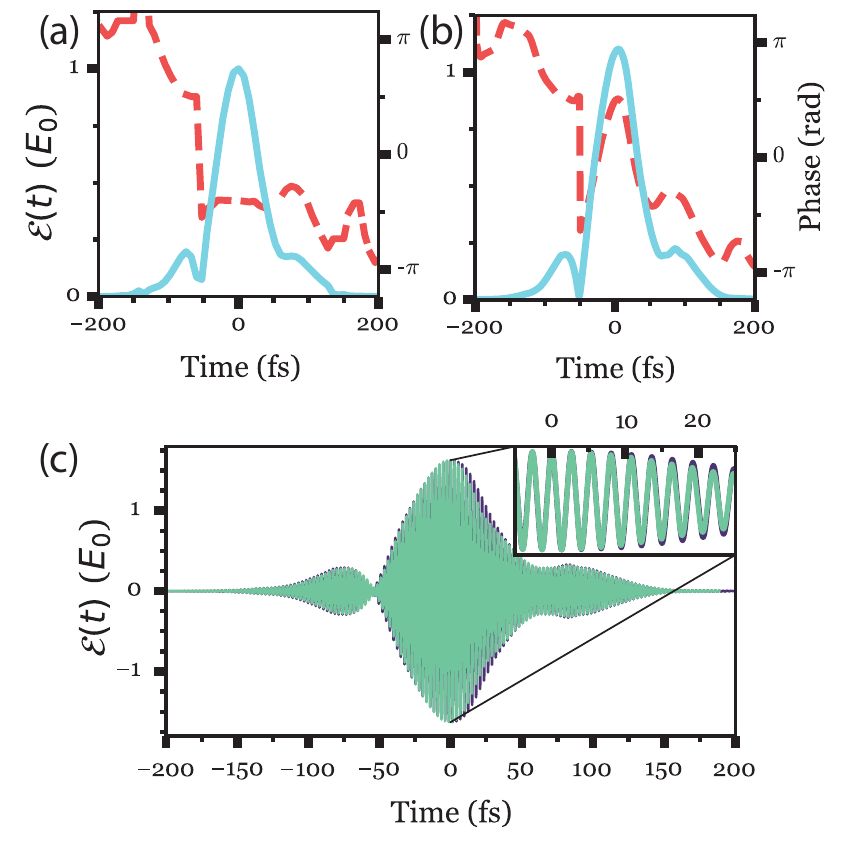}
\caption{(a)Amplitude (solid) and phase (dashed) of the electric field before the DLA, retrieved from via FROG. The pulse consists of a short peak of FWHM 45~fs with satellite lobes caused by residual higher-order(3+) dispersion. (b) Amplitude(solid) and phase(dash) of an $E_0=4.75$~GV/m electric field after simulated propagation through 499~$\mu$m of fused silica. (c) Comparison of the waveform input to the FDTD simulation (dark purple) and of the (longitudinal) waveform evaluated at the position marked by a star in Fig.~\ref{fig:setup} (light green). The input waveform has been scaled in amplitude and offset in phase to aid the comparison.}
\label{fig:simFROG}
\end{figure}

The effect of the nonlinear material response on the DLA energy gain is simulated in three steps (Fig.~\ref{fig:simFROG}): first, the incident amplitude and phase (up to a time-reversal ambiguity) are reconstructed from the FROG image of the laser pulse measured before the DLA 
; second, that beam envelope is propagated through 499~$\mu$m of glass by solving a generalized nonlinear Schr\"{o}dinger equation (NLSE) (see supplementary material) using a split-step Fourier solver on an adaptive grid~\cite{grace_gaffe_2009}; and finally the grating layer is simulated by the commercial FDTD code Lumerical,~\cite{noauthor_fdtd_nodate} using the output of the NLSE as a source.

For incident fields near $E_0\sim5$~GV/m the NLSE predicts significant nonlinear contributions to the effective length of the accelerator. The propagator includes effects such as self-focusing, self-steepening, Raman scattering~\cite{agrawal_chapter_2013-1} and multiphoton absorption~\cite{hoyo_rapid_2015}, but in Fig.~\ref{fig:simFROG}(b) the dominant feature is the intensity-dependent phase modulation. Self-focusing is evident in the increased value of $\mathcal{E}(t)$, while the effects of self-steepening and multiphoton absorption are still quite subtle. Note that free carrier generation is not included in the propagator, because a post-hoc calculation of the free carrier density~\cite{sudrie_femtosecond_2002, schiffrin_optical-field-induced_2012} suggests that the induced phase change is negligible compared to the Kerr effect until very near the damage threshold. 

One important result from the chain of simulations is the observation that, independent of laser intensities we studied, the grating layer preserves the complex amplitude and phase of the input pulse (up to a scale factor in amplitude and an offset in phase). This is shown in Fig.~\ref{fig:simFROG}(c) by comparing the waveform input at the grating layer to the waveform evaluated in the center of the vacuum gap (marked by a star in Fig.~\ref{fig:setup}). The input and output pulses are nearly identical except for a small delayed reflection. This is a consequence of the fact that the bandwidth of the laser remains smaller than the bandwidth of the accelerator~\cite{soong_electron_2014}. Thus, it is a good approximation to bypass the computationally intensive simulation of the grating layer and directly use the pulse envelope at the entrance of the grating layer to track particles, as in Eq.~\ref{eq:1}, with the calculated $\alpha$~$=$~$0.18-0.23$ depending on the relative longitudinal alignment of the two grating layers.

The variation of the field envelope input to the grating layer as a function of intensity are the cause of the saturation observed in Fig.~\ref{fig:gradvse0}. This effect can be pre-compensated by applying anomalous dispersion (i.e. stretching the pulse to length $\tau$) before the DLA so that the Kerr effect `compresses' the dispersed pulse to have a flat phase. Without the nonlinear mechanism, the dispersion lengthens the pulse, but it also introduces a chirp which causes electrons to dephase, keeping the energy gain constant (i.e. independent of dispersion). When the nonlinear propagation is included, the pulse has a lower field ($E\propto \sqrt{\frac{1}{\tau}}$), but a longer interaction length ($L_\text{eff}\propto\tau$) is responsible for a net energy gain increase ($\propto$ $\sqrt{\tau}$). We can illustrate this process by re-writing Eq.~\ref{eq:1} in the Fourier domain:
\begin{equation}\label{eq:2}
\Delta E=\left| \int_{-\infty}^{\infty}\alpha E_0 \mathcal{F} \left(\mathcal{E}(\vec{k},\omega)\right)[\vec{r}=\vec{r_0}+\vec{\beta}ct,t]dz \right|
\end{equation}
where $\mathcal{E}(\vec{k},\omega)$ is the accelerating field, $E_z$, written in the Fourier domain, and $\mathcal{F}$ is the Fourier transform operator. By exchanging the order of integration we find a Dirac delta function enforcing the plane-wave phase matching condition ($\delta\left(k_g (\beta ct)-\omega t\right)$) and we arrive at simple expression for the energy gain in terms of the laser spectrum:
\begin{equation}\label{eq:3}
\Delta E \propto \alpha E_0 \left| \mathcal{E}(\omega=0)\right|
\end{equation}
which shows that the energy gain is proportional to the phase-matched ($\omega_0$) frequency component of the laser electric field. Note that we have ignored the angular distribution of the drive laser wavenumbers $\vec{k}$, which results in blurring the Dirac delta function to an $\approx$1~nm bandwidth. From this formulation, it follows that dispersion---which does not alter the bandwidth of the laser---will not change the energy gain of the DLA by itself. Self-phase modulation, however, changes the bandwidth of the pulse by applying a nonlinear phase in the time-domain so that the initial dispersion becomes a sensitive parameter for a high intensity drive laser.

\begin{figure}[h]
\centering
\includegraphics[width=.45\textwidth]{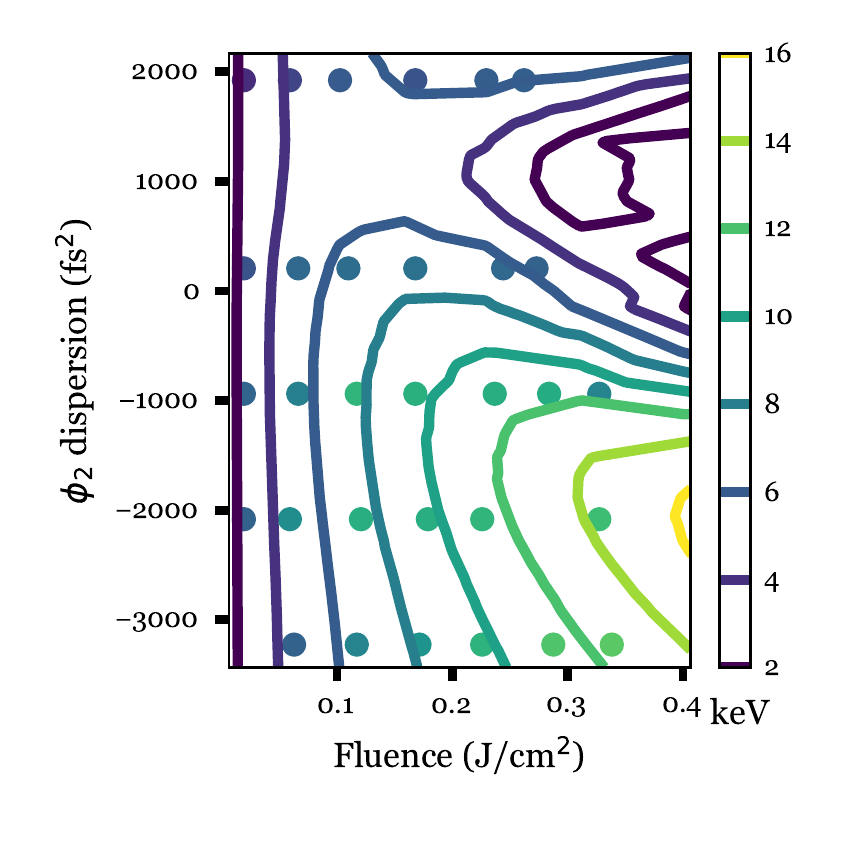}
\caption{The contours illustrate a theoretical calculation of the energy gain experienced by the electron beam as a function of fluence and dispersion (as determined at the entrance to the fused silica wafer). The scatter plot illustrates an experimental measurement of the energy gain of an electron beam interacting with such a drive laser.}
\label{fig:comp_nonlin}
\end{figure}

Such behavior is well reproduced in the experimental measurements and fully supported by the pulse propagation and particle tracking calculations (see Fig.~\ref{fig:comp_nonlin}). For low pulse energy, decreasing the dispersion has no effect; while for high pulse energy, decreasing the dispersion results in a rapid increase in energy gain. Experimentally, the second order dispersion is controlled by adjusting the spacing of the compressor gratings after the regenerative amplifier. The introduced dispersion can be directly calculated~\cite{treacy_optical_1969} and separately calibrated by offline FROG measurements, while the optimum compression (or zero dispersion) point is determined by second harmonic generation in a thin BBO crystal located in the same plane as the DLA, as in Fig.~\ref{fig:gradvse0}. The data here are recorded using a DLA with an 800~nm vacuum gap (having smaller $\alpha$ and larger transmission) in order to increase the signal-to-noise ratio from the energy spectra. As in Fig.~\ref{fig:gradvse0}, the differences between the simulation and measured data can be attributed to uncertainty in the electron beam parameters and alignment of the laser. Nonetheless, the strong agreement over a wide range of parameters in the fluence/dispersion scan demonstrates the importance of pulse shaping for accelerator optimization.

In conclusion, the fused silica, dual-grating, DLA structure is shown to produce 1.8~GV/m accelerating fields when driven by a 9~GV/m, 45~fs laser. Taking advantage of such an intense power source requires elimination of the intensity-dependent phase produced by the Kerr effect, either by using thinner substrates or by pre-compensating with anomalous dispersion. We have demonstrated the effectiveness of shaping the pulse profile for keeping electrons synchronous with the accelerating wave over many periods. Our work highlights the unique sensitivity of the DLA to pulse shaping of the drive laser: where other laser accelerators, such as laser plasma wakefield acceleration, are sensitive only to the pulse envelope; the DLA is sensitive to both the intensity and phase. This offers an extra degree of freedom in extending the acceleration region and controlling the beam dynamics over significant lengths. For example, more complex pulse shaping could be used to create alternate gradient focusing~\cite{rosenzweig_transverse_1994,naranjo_stable_2012,xie_laser_1998}, or as a means of changing the resonant velocity.

This work was supported by the U.S. Department of Energy, Office of Science, under Contract no. DE-AC02-76SF00515, and by the Gordon and Betty Moore Foundation under grant GBMF4744 (Accelerator on a Chip).

\bibliographystyle{apsrev4-1}
\bibliography{zotero_update}

\clearpage
\pagebreak
\section{Supplemental Material: Nonlinear propagation in bulk fused silica}
We use numerical laser propagation to relate the experimentally measured laser properties to the fields inside the DLA. Where possible we avoid lengthy electromagnetic simulations by taking the slowly varying envelope approach. We describe a pulse propagating in the $y$ direction as as $A e^{i(k_0 y- \omega_0 t)}$, where $A$, the envelope, can take on complex values, but is assumed to vary slowly relative to the optical fields. In a nonlinear medium this approach leads to a generalized nonlinear Schr\"{o}dinger equation (NLSE)~\cite{agrawal_chapter_2013-1,hoyo_rapid_2015}

\begin{equation}\label{eq:A1}
\frac{\partial A}{\partial{y}}=\left[\hat{D_f}+\hat{D_s}+\hat{N}\right]
\end{equation}
where $\hat{D_f},\hat{D_f},\hat{N}$ are the diffraction, dispersion, and nonlinear operators, respectively:
\begin{align}\label{eq:A2}
\hat{D_f}=&\left(\frac{i \lambda}{4 \pi n_0} \right)\left(1-\frac{i \lambda}{2\pi c}\partial_t\right)\left(\partial_x^2+\partial_z^2\right)\\
\hat{D_s}=&\left(\frac{-i}{2}\right)k^{(2)}\partial_t^2-\frac{1}{2}\alpha\\
\hat{N}=&\left(\frac{i2\pi n_2}{\lambda}\right)\left|A^2\right|-\frac{1}{A}\left(\frac{n_2}{c}\right)\partial_t\left(A\left|A^2\right|\right) \\ &-\frac{i2\pi n_2}{\lambda}T_r\partial_t\left|A^2\right|-\frac{\beta_6}{2}\left|A^{10}\right|
\end{align}

Note that the equations are written in the moving frame, $t=t_0-y/v_g$, and include dispersion, $k^{(2)}$, absorption, $\alpha$, six photon absorption, $\beta_6$, and third order polarization, $n_2$, with a linearized Raman response, $T_r$. 

Given an initial field profile $A\left(x,z,y=0\right)$, the field after 500~$\mu$m of glass is calculated via a finite difference approach to Eq.~\ref{eq:A1}. The propagation is implemented using the `Generalised Adaptive Fast-Fourier Evolver' (GAFFE)~\cite{grace_gaffe_2009} which uses a split-step Fourier solver and an adaptive grid in order to rapidly compute the right hand side of Eq.~\ref{eq:A1} without aliasing.  

The model inputs for these simulations are cataloged in Table \ref{tab:1}. Note that the 6-photon cross section may be calculated from the Keldysh formula in the low intensity limit or fit to give an effective value at intermediate intensities~\cite{sudrie_femtosecond_2002,hoyo_rapid_2015}.

For the relevant initial conditions we find that Kerr self-phase-modulation is the primary nonlinear phenomena, followed by self-focusing and then 6-photon absorption and self-steepening. Fig.~\ref{fig:simFROG}(b) of the main text shows the propagation of the envelope in Fig.~\ref{fig:simFROG}(a) through 499~$\mu$m of fused silica for a peak field of 4.75~GV/m. Self-phase modulation is evident in that the phase (dotted) follows the intensity (the square of the illustrated electric field). Self-focusing is also evident: the peak of the envelope in Fig.\ref{fig:simFROG}(b) exceeds that of Fig.~\ref{fig:simFROG}(a). Self-steepening is barely visible, but begins to present itself more prominently as $E_0$ increases. Similarly the 6-photon absorption begins to erode energy from the peak which alters the onset of  filimentation in the pulse.  

The generation of free carriers~\cite{sudrie_femtosecond_2002} is neglected in eq.(\ref{eq:A1}) because post-hoc estimates suggest that $\Delta n$ due to free carriers is a small fraction of the $\Delta n$ due to self phase modulation. At $E_0$=5~GV/m---where self-phase modulation saturates the DLA interaction ($\Delta E$)---multi-photon absorption is still the dominate mechanism (the Keldysh parameter is $\sim$3) and the maximum electron density is estimated to be less than 1$\cdot$$10^{15}$cm$^{-3}$, which would cause a change in $n$ 0.025\% that of $n_2 I$. For $E_0>10$~GeV/m the free-carrier population begins to become significant (but now increasing as $\sim I^4$), which likely contributes to the observed damage threshold around that point.

\begin{table}\label{tab:1}
	\centering
	\caption{Table of model inputs}\label{tab:1}
		\begin{tabular}{|l|c|r|}
		\hline
			Parameter & value & notes\\ 	\hline
			\hline
			Material &  & \\ \hline		
			$n_0$ & 1.45 & \\ \hline
			$n_2$ & 2.45$\cdot$10$^{16}~$cm$^2$/W & \cite{milam_review_1998} \\  \hline
			$k^{(2)}$ & 36.163~fs$^2$/mm & \\ \hline
			$\alpha$ & 10$\cdot 10^{-6}$~1/cm & \\ \hline
			$\beta_6$ & 5$\cdot 10^{-83}$~m$^9$/W$^6$ & \cite{hoyo_rapid_2015}  \\ \hline
			$\tau_r$ & 3~fs & \cite{agrawal_chapter_2013-1} \\ \hline
			\hline
			Laser & & \\ \hline
			$\lambda$ & 800~nm & \\ \hline
			$w_x$ & 45~$\mu$m & Gaussian \\ \hline
			$w_z$ & 500~$\mu$m & Gaussian \\ \hline
			$\tau$ & $\sim$45~fs & FROG trace \\ \hline
			Energy & $2-200$~$\mu$J &  \\ \hline			
		\end{tabular}
\end{table}

\end{document}